# The Effect of Phasor Measurement Units on the Accuracy of the Network Estimated Variables

H. Abdollahzadeh Sangrody
Power and Water University of Technology
Tehran, Iran
h.sangrody@gmail.com

M.T. Ameli
Power and Water University of Technology
Tehran, Iran
ameli@pwut.ac.ir

M.R.Meshkatoddini
Power and Water University of Technology
Tehran, Iran
meshkatoddini@pwut.ac.ir

***Abstract*— The most commonly used weighted least square state estimator in power industry is nonlinear and formulated by using conventional measurements such as line flow and injection measurements. PMUs (Phasor Measurement Units) are gradually adding them to improve the state estimation process. In this paper the way of corporation the PMU data to the conventional measurements and a linear formulation of the state estimation using only PMU measured data are investigated. Six cases are tested while gradually increasing the number of PMUs which are added to the measurement set and the effect of PMUs on the accuracy of variables are illustrated and compared by applying them on IEEE 14, 30 test systems.**

## I. INTRODUCTION

State estimation is a key element of the online security analysis function in modern power system energy control centers. The function of state estimation is to process a set of redundant measurements to obtain the best estimate of the current state of a power system. State estimation is traditionally solved by the weighted least square algorithm with conventional measurements such as voltage magnitude, real and reactive power injection, real and reactive power flow [1]. Recently, synchronized phasor measurement techniques based on a time signal of the GPS (Global Positioning System) are introduced in the field of power systems. A PMU, when placed at a bus, can measure the voltage phasor at the bus, as well as the current phasors through the lines incident to the bus. It samples the ac voltage and current waveforms while synchronizing the sampling instants with a GPS clock. The computed values of voltage and current phasors are then time-stamped and transmitted by the PMUs to the local or remote receiver [2]-[4]. The traditional state estimation is by nature a nonlinear problem. The most commonly used approach is Weighted Least Squares which converts the nonlinear equations into the normal equations by using first-order Taylor series. However, the state estimation equations for PMU measurements are inherently linear equations. Some research has been conducted to try to formulate the mixed set of traditional and PMU measurements. The natural approach is to treat PMU measurements as additional measurements to be appended to traditional measurements, which causes the additional computation burden of calculation. Another approach is to use the distributed scheme for the mixed state estimation [5-9]. The problem of finding optimal PMU locations for power system state estimation is well investigated in the literature [10]–[13]. This paper shows the effect of PMUs on the accuracy of the estimated variables. Six cases are tested by gradually increasing the PMU numbers and applying them on IEEE 14, 30 test systems. In the first case, state estimation without any PMU and in the sixth case, linear formulation of the state estimation using only PMU measured data is discussed. In the other four cases, hybrid state estimation with different number of added PMUs to the conventional measurement set are tested.

## II. WEIGHTED LEAST SQUARED STATE ESTIMATION METHOD

As shown “(1)”, this method minimizes the weighted sum of squares of the residuals

$$J(x)=\sum_{i=1}^{N_m}\frac{(z_i-h_i(x))^2}{\delta_i{}^2}=\left[z-h(x)\right]^T R^{-1}\left[z-h(x)\right]. \quad (1)$$

Where in this equation $z$ is measurement vector, $x$ state vector, $\sigma$ standard deviation and $h$ is the nonlinear function relating measurement $i$ to the state vector $x$. $R$ is measurement covariance matrix is given by “(2)”.

$$R = diag\left[\delta_1^2, \delta_2^2, \ldots, \delta_{Nm}^2\right] \ . \tag{2}$$

At the minimum value of the objective function, the first-order optimality conditions have to be satisfied. These can be expressed in compact form as follows:

$$g(x) = -\frac{\partial J(x)}{\partial x} = -H^T(x)R^{-1}(z - h(x)) = 0\,. \tag{3}$$

Where

$$H(x) = \frac{\partial h(x)}{\partial x}\,. \tag{4}$$

The nonlinear function $g(x)$ can be expanded into its Taylor series around the state vector $x^k$ neglecting the higher order terms. An iterative solution scheme known as the Gauss-Newton method is used to solve "(3)":

$$x^{k+1} = x^k - \left[G(x^k)\right]^{-1} . g(x^k) \tag{5}$$

Where, k is the iteration index and $x^k$ is the solution vector at iteration k. $G(x^k)$ is called the gain matrix, and expressed by:

$$G(x^k) = \frac{\partial g(x^k)}{\partial x} = H^T(x^k)R^{-1}H(x^k) \tag{6}$$

$$g(x^k) = -H^T(x^k)R^{-1}\left[z - h(x^k)\right] \tag{7}$$

These iterations are going on until the maximum variable difference satisfies the condition, $'Max\left|\Delta x^k\right| < \varepsilon'$. Consider a system having (*N*) buses; the state vector will have (*2N-1*) components which are composed of (*N*) bus voltage magnitudes and (*N-1*) phase angles.

## III. Conventional State Estimation

There are three most commonly used measurement types in conventional state estimation. They are bus power injections, line power flows and bus voltage magnitudes. These measurement equations can be expressed using the state variables. Jacobian matrix *H* has rows at each measurement and columns at each variable. Considering "(4)", power injection and power flow in Fig. 1, *H* matrix components corresponding to these measurements are partial derivation of each variable [14].

$$H = \begin{bmatrix} \dfrac{\partial P_i}{\partial \delta} & \dfrac{\partial P_i}{\partial V} \\ \dfrac{\partial P_{ij}}{\partial \delta} & \dfrac{\partial P_{ij}}{\partial V} \\ \dfrac{\partial Q_i}{\partial \delta} & \dfrac{\partial Q_i}{\partial V} \\ \dfrac{\partial Q_{ij}}{\partial \delta} & \dfrac{\partial Q_{ij}}{\partial V} \\ 0 & \dfrac{\partial V_{Mag}}{\partial V} \end{bmatrix} \tag{8}$$

In this matrix $\delta$ and $V$ are state variables, $P_i$ and $Q_i$ are real and reactive power injection at bus *i*. $P_{ij}$, $Q_{ij}$ are real and reactive power flow from bus *i* to bus *j*. In this condition of estimation, relation between measurement data and state variables are nonlinear and its final solution depend on an iterative solution scheme expressed "(5)".

## IV. Hybrid State Estimation

One PMU can measure the voltage and the current phasors. The equivalent $\pi$ model of the line connecting buses *i* and *j* with assumption of a PMU connected to bus *i* is shown in Fig. 1.

If $Y_{ij} = g_{ij} + jb_{ij}$ is defined as the series admittance and $Y_{si} = g_{si} + jb_{si}$ as the shunt admittance, current phasor measurements can be written in rectangular coordinates as shown in Fig. 1. The expressions for $C_{ij}$ and $D_{ij}$ are:

$$\begin{aligned} C_{ij} = &\left|V_i Y_{si}\right|\cos(\delta_i + \theta_{si}) - \\ &\left|V_j Y_{ij}\right|\cos(\delta_j + \theta_{ij}) + \left|V_i Y_{ij}\right|\cos(\delta_j + \theta_{ij}) \end{aligned} \tag{9}$$

$$\begin{aligned} D_{ij} = &\left|V_i Y_{si}\right|\sin(\delta_i + \theta_{si}) - \\ &\left|V_j Y_{ij}\right|\sin(\delta_j + \theta_{ij}) + \left|V_i Y_{ij}\right|\sin(\delta_j + \theta_{ij}) \end{aligned} \tag{10}$$

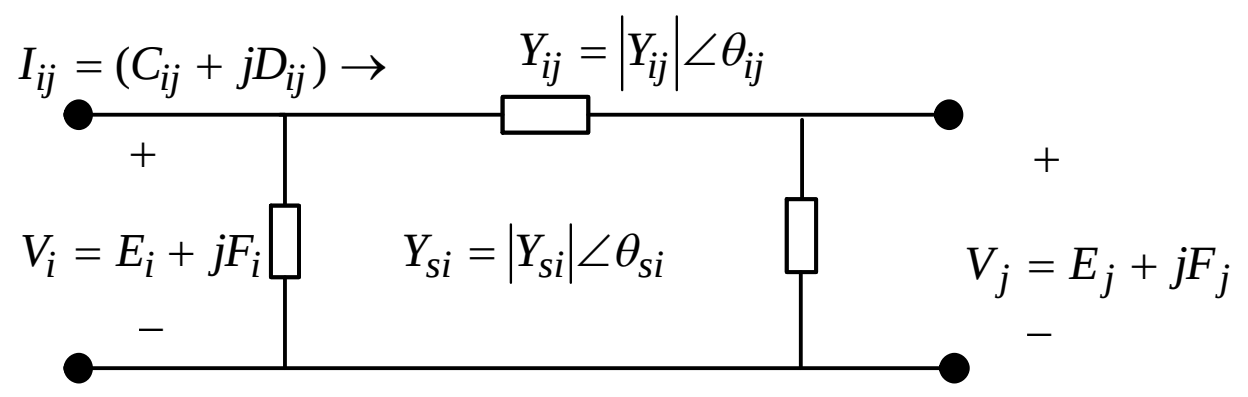

Figure 1. Transmission Line Model

The entries of the measurement Jacobian $H$ corresponding to the real and reactive parts of the current phasors are:

$$\frac{\partial C_{ij}}{\partial V_i} = |Y_{si}|\cos(\delta_i + \theta_{si}) + |Y_{ij}|\cos(\delta_i + \theta_{ij}) \quad (11)$$

$$\frac{\partial C_{ij}}{\partial V_j} = -|Y_{ij}|\cos(\delta_j + \theta_{ij}) \quad (12)$$

$$\frac{\partial C_{ij}}{\partial \delta_i} = -|V_i Y_{si}|\sin(\delta_i + \theta_{si}) - |V_i Y_{ij}|\sin(\delta_i + \theta_{ij}) \quad (13)$$

$$\frac{\partial C_{ij}}{\partial \delta_j} = |V_j Y_{ij}|\sin(\delta_j + \theta_{ij}) \quad (14)$$

$$\frac{\partial D_{ij}}{\partial V_i} = |Y_{si}|\sin(\delta_i + \theta_{si}) + |Y_{ij}|\sin(\delta_i + \theta_{ij}) \quad (15)$$

$$\frac{\partial D_{ij}}{\partial V_j} = -|Y_{ij}|\sin(\delta_j + \theta_{ij}) \quad (16)$$

$$\frac{\partial D_{ij}}{\partial \delta_i} = |V_i Y_{si}|\cos(\delta_i + \theta_{si}) + |V_i Y_{ij}|\cos(\delta_i + \theta_{ij}) \quad (17)$$

$$\frac{\partial C_{ij}}{\partial \delta_j} = -|V_j Y_{ij}|\cos(\delta_j + \theta_{ij}) \quad (18)$$

The measurement vector z contains $\delta$ , $C_{ij}$ and $D_{ij}$ as well as the power injections, power flows and voltage magnitude measurements.

$$Z = \left[ P_{inj}^T, Q_{inj}^T, P_{flow}^T, Q_{flow}^T, |V|^T, \delta^T, C_{ij}^T, D_{ij}^T \right] \quad (19)$$

Generally, those measurements received from PMUs are more accurate with small variances compared to the variances of conventional measurements. Therefore, including PMU measurements is expected to produce more accurate estimates.

## V. LINEAR FORMULATION OF STATE ESTIMATION USING ONLY PMUS

If the measurement set is composed of only voltage and current measured by PMUs, the state estimation can be formulated as a linear problem. The state vector and measurement data can be expressed in rectangular coordinate system. As shown in Fig. 1 a PMU located at bus $i$ measured voltage $V_i$ and line current $I_{ij}$ . The voltage measurement ( $V_i = |V_i|\angle\delta_i$ ) can be expressed as ( $V_i = E_i + jF_i$ ), and the current measurement can be expressed as ( $I_{ij} = C_{ij} + jD_{ij}$ ). In this condition of estimation, measurement vector $z$ and sate vector $x$ are:

$$z = \left[E_i \ C_{ij} \ F_i \ Di_j\right]^T \quad (20)$$

$$x = \left[E_i \ E_j \ F_i \ F_j\right]^T \quad (21)$$

In Fig. 1, line current flow $I_{ij}$ can be expressed as a linear function of voltages.

$$\begin{aligned} C_{ij} + jD_{ij} = &\left[(g_{ij} + jb_{ij}) + (g_{si} + jb_{si})\right] \ (E_i + jF_i) \\ &-(g_{ij} + jb_{ij})(E_j + jF_j) \end{aligned} \quad (22)$$

Jacobian matrix $H$ components are expressed by

$$\frac{\partial C_{ij}}{\partial E_i} = g_{ij} + g_{si} \quad (23)$$

$$\frac{\partial C_{ij}}{\partial E_j} = -g_{ij} \quad (24)$$

$$\frac{\partial C_{ij}}{\partial F_i} = -b_{ij} - b_{si} \quad (25)$$

$$\frac{\partial C_{ij}}{\partial F_j} = b_{ij} \quad (26)$$

$$\frac{\partial D_{ij}}{\partial E_i} = b_{ij} + b_{si} \quad (27)$$

$$\frac{\partial D_{ij}}{\partial E_j} = -b_{ij} \quad (28)$$

$$\frac{\partial D_{ij}}{\partial F_i} = g_{ij} + g_{si} \quad (29)$$

$$\frac{\partial D_{ij}}{\partial F_j} = -g_{ij} \tag{30}$$

Then, the estimated value $\hat{x} = \hat{E}_i + j\hat{F}_i$ can be obtained by solving the linear equation below:

$$\hat{x} = (H^T R^{-1} H)^{-1} H^T R^{-1} \hat{z} \ . \tag{31}$$

This is very simple and fast, because it doesn't need any iteration. In addition covariance matrix $R$ in "(31)" is very smaller than covariance matrix of conventional measurement, so the estimated variables are very accurate.

## VI. SIMULATION RESULTS

To investigate the effect of PMUs on the accuracy of estimated variables, several cases are tested with different number of added PMUs to the conventional measurement set. Two different IEEE test systems (IEEE 14, IEEE 30 bus system) are tested with 6 different cases which are shown in table I. Fig. 2 and Fig. 3 show the network diagrams for each system. Arrow with circle at the bus means a pair of real and reactive power injection measurements and any point on the transmission line; it means a pair of real and reactive power flow measurements.

TABLE I. SIX DIFFERENT CASES BY ADDING PMUS

| | |
|---|---|
| **Case 1** | Conventional Measurements with No PMUs |
| **Case 2** | Conventional Measurements with PMUs of (10% of bus number) |
| **Case 3** | Conventional Measurements with PMUs of (20% of bus number) |
| **Case 4** | Conventional Measurements with PMUs of (30% of bus number) |
| **Case 5** | Conventional Measurements with PMUs of (40% of bus number) |
| **Case 6** | Only Minimum PMUs |

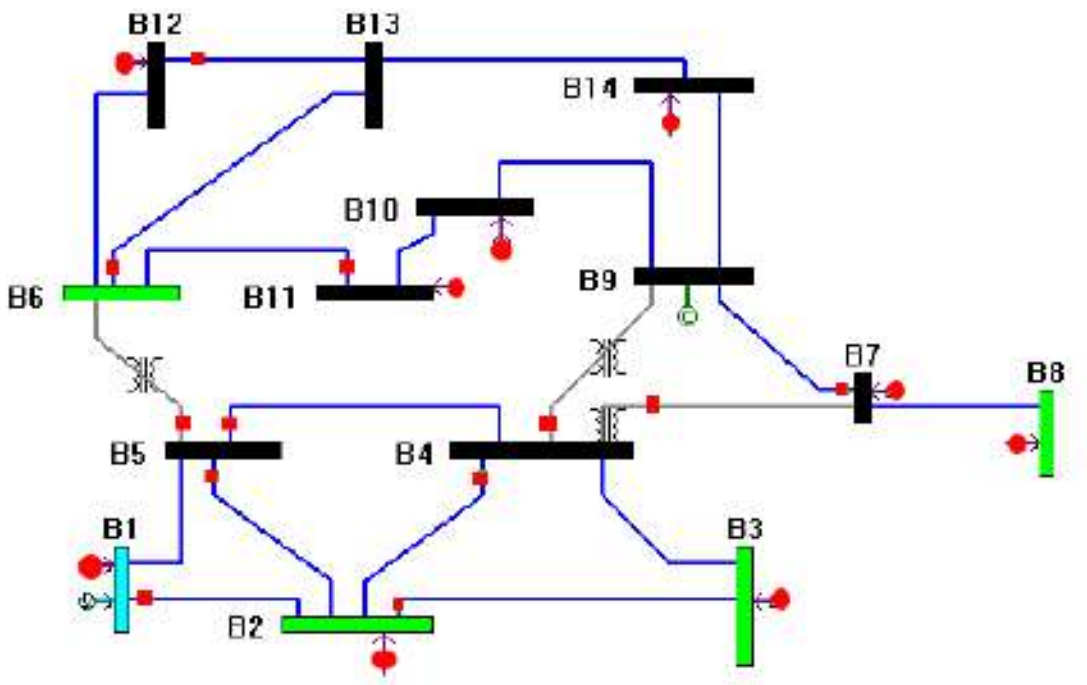

Figure 2. IEEE14 Bus System Diagram with Conventional Measurements

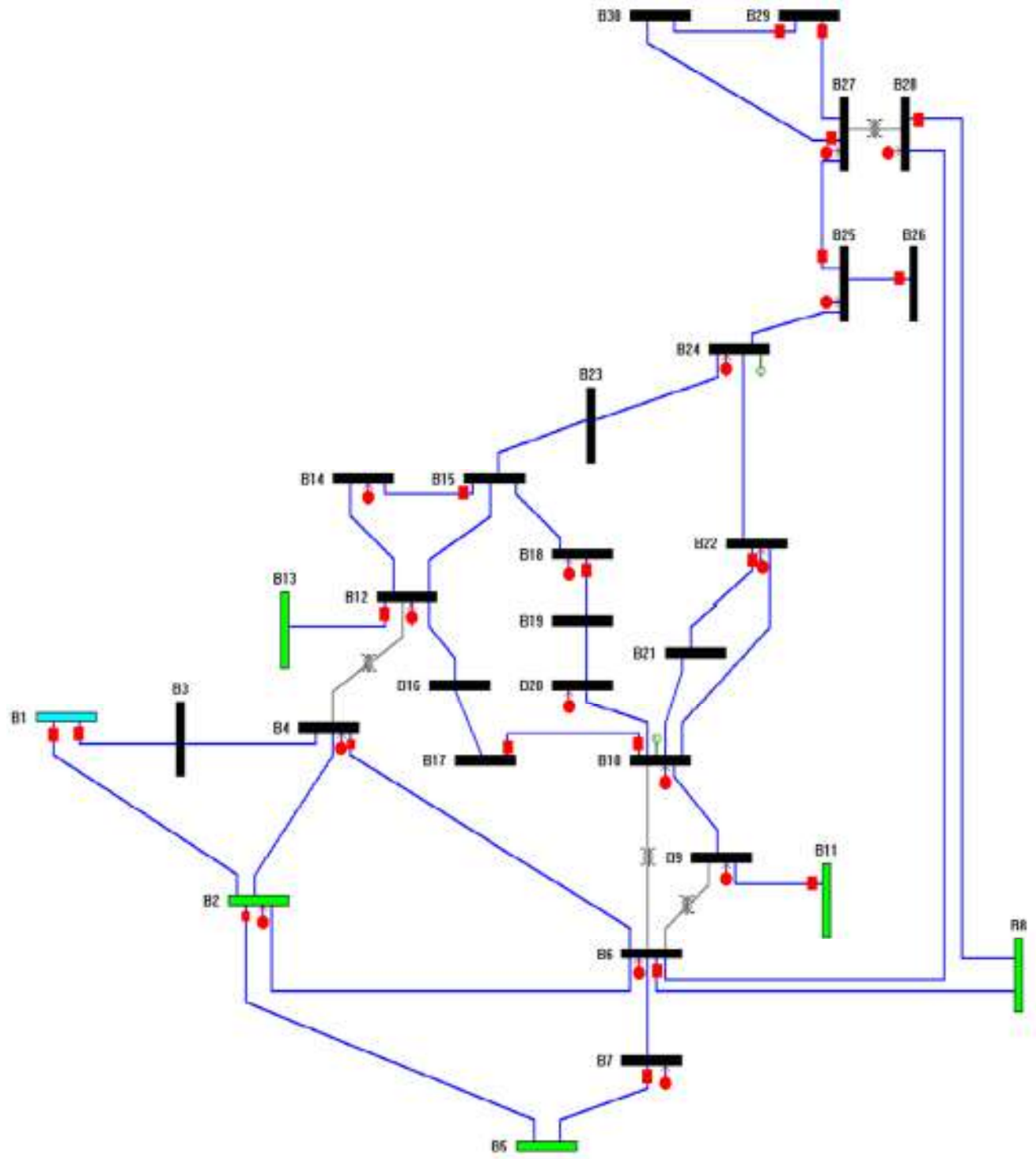

Figure 3. IEEE30 Bus System Diagram with Conventional Measurements

Each network has a voltage magnitude measurement connected to bus 1. Table II has more detailed information about the conventional measurement set installed in the networks.

The setting of error standard deviations for power injection, power flow and voltage magnitude are 0.01, 0.08 and 0.04 respectively. A PMU has much smaller error deviations than other conventional measurements as 0.00001. PMUs are located at buses 2, 6, 7 and 9 in IEEE 14 bus system, and at buses 3, 5, 6, 9, 10, 12, 19, 23, 25 and 29 in IEEE 30 bus system at case 6 [10]. One of the ways of representing the level of state estimation accuracy is to refer the covariance of the estimated variables. The covariance of the estimated variable vector is obtained from the inverse diagonal elements of gain matrix. The accuracy of two variables (voltage magnitude, voltage angle) is investigated separately. Fig. 4 and Fig. 5 show the accuracy of the estimated voltage magnitudes of each system. Fig. 6 and Fig .7 show the accuracy of the estimated voltage angles of each system.

TABLE II. VARIABLE NUMBERS, MEASUREMENT TYPE AND NUMBERS

| **Network** | **Variables** | **Measurements** | | | |
|---|---|---|---|---|---|
| | | ***Power injection*** | ***Power flow*** | ***Voltage magnitude*** | ***Total*** |
| **IEEE 14 bus system** | 27 | 18 | 24 | 1 | 43 |
| **IEEE 30 bus system** | 59 | 38 | 56 | 1 | 95 |

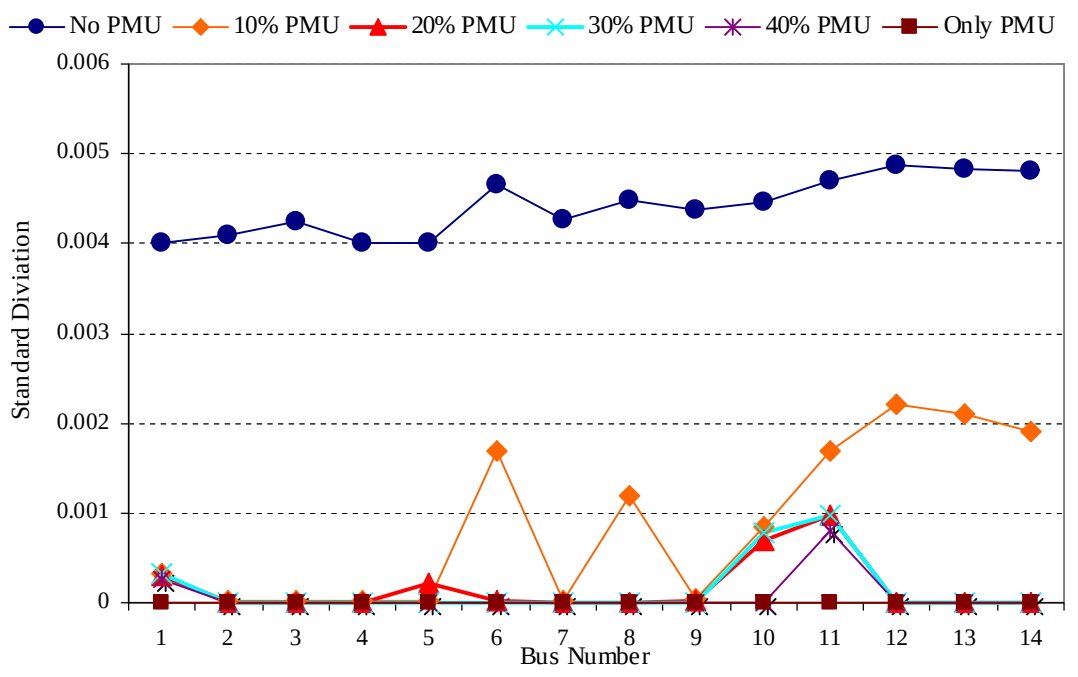

Figure 4. Accuracy of |V| of IEEE14 Bus System with PMUs

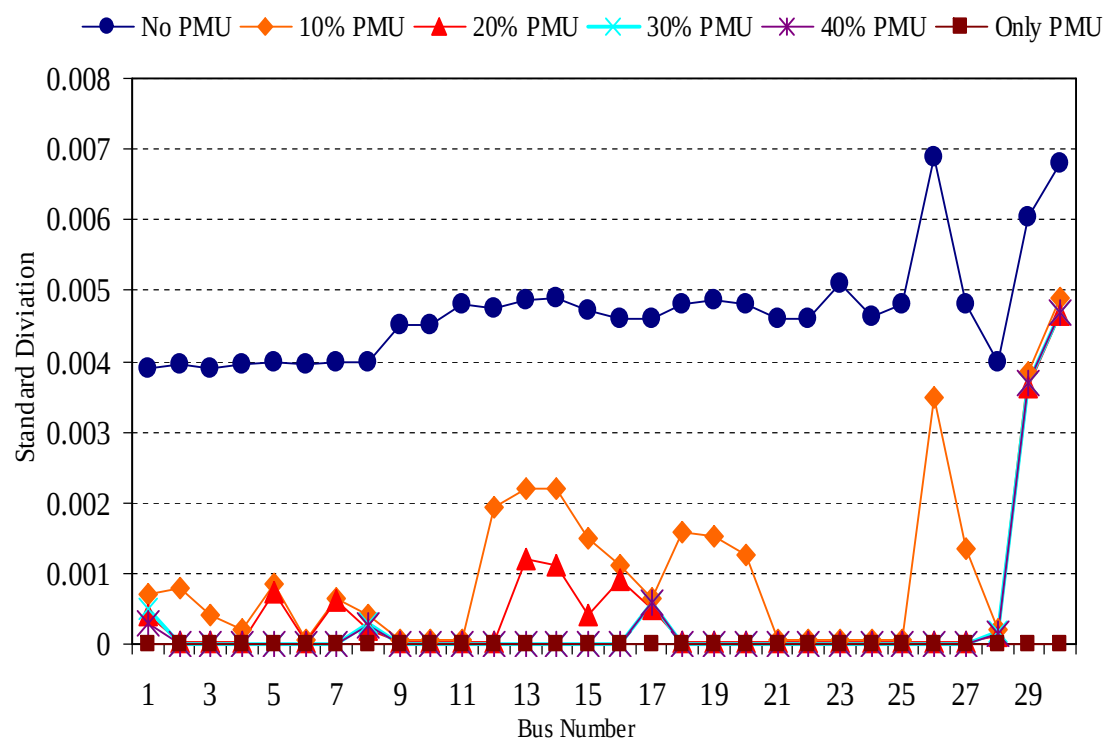

Figure 5. Accuracy of |V| of IEEE30 Bus System with PMUs

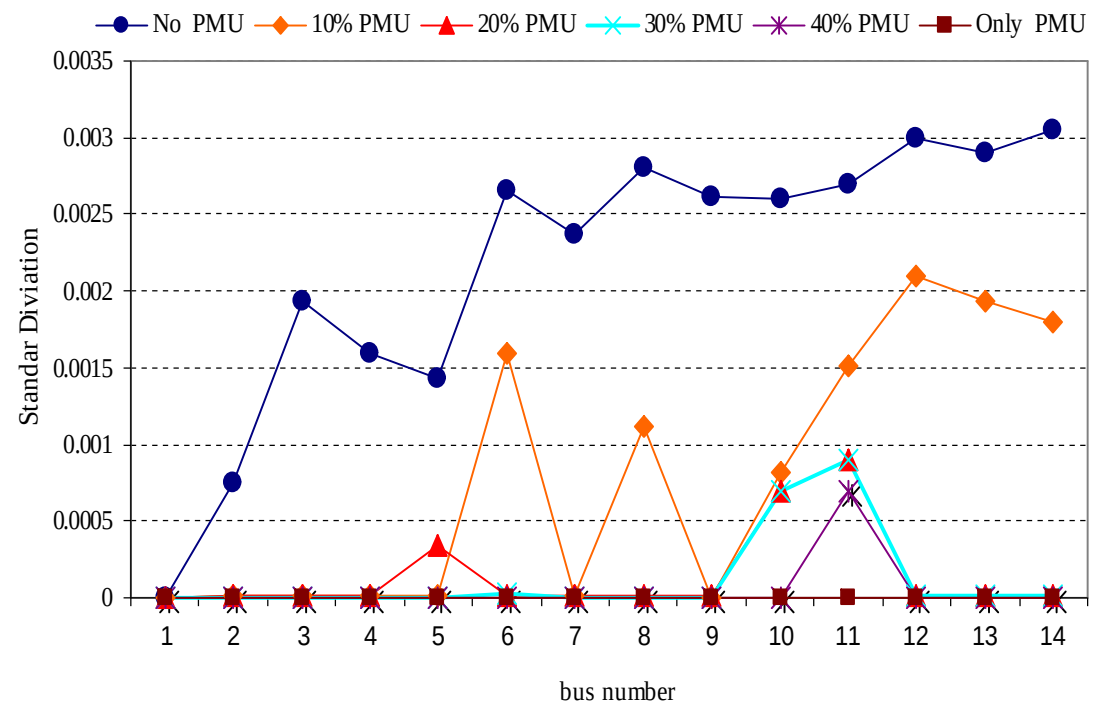

Figure 6. Voltage Angle Accuracy of IEEE14 Bus System with PMUs

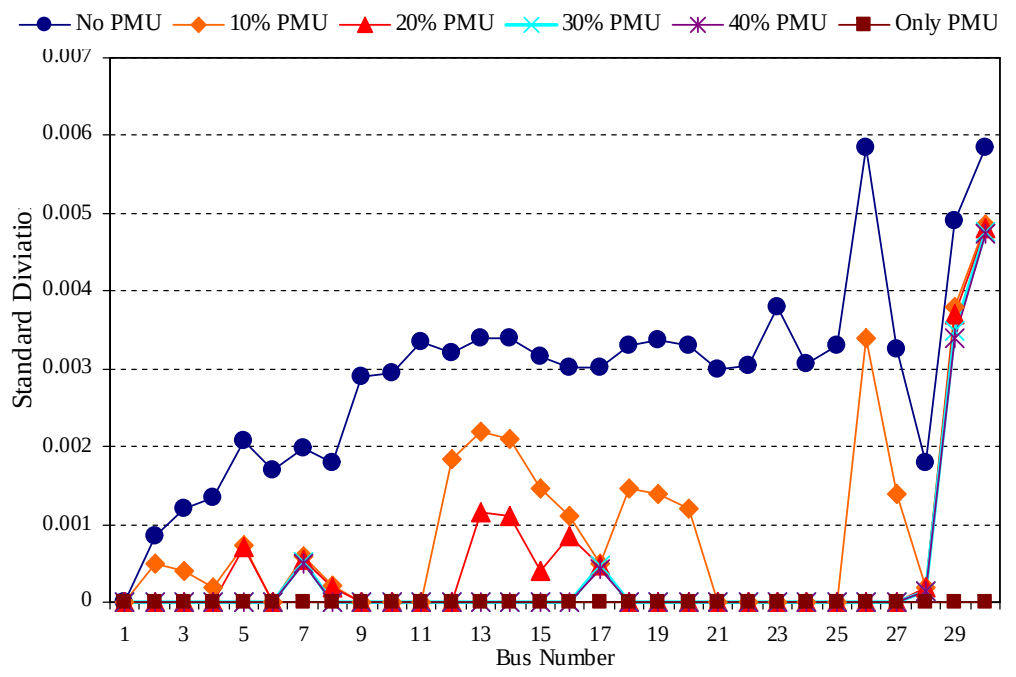

Figure 7. Voltage Angle Accuracy of IEEE30 Bus System with PMUs

These figures show the effect of the PMUs on the accuracy of the estimated variable. Average valued S.D. (Standard Deviation) of each variable and their percentage values are shown in tables III and IV. The percentage values in these tables mean that how the S.D. values at each cases are decreased compared to the S.D. of 'Case 1' which is forced to be set as 100%. In case of 'Only PMUs', it becomes nearly zero. Average valued S.D. of voltage magnitude and voltage angle are shown in Fig. 8 and Fig. 9 respectively.

TABLE III. AVERAGE ERROR STANDARD DEVIATIONS OF THE VOLTAGE MAGNITUDE

| cases | IEEE 14 Bus | | IEEE 30 Bus | |
|---|---|---|---|---|
| | ***Average Error S.D.*** | ***Percentage*** | ***Average Error S.D.*** | ***Percentage*** |
| **No PMUs** | 0.0043741 | 100% | 0.0046548 | 100% |
| **10% PMUs** | 0.0008449 | 19.31% | 0.0010444 | 22.43% |
| **20% PMUs** | 0.0001727 | 3.95% | 0.0004986 | 10.70% |
| **30% PMUs** | 0.0001497 | 3.42% | 0.0003240 | 6.96% |
| **40% PMUs** | 0.0000810 | 1.85% | 0.0003229 | 6.94% |
| **Only PMUs** | 0.0000055 | 0.13% | 0.0000042 | 0.09% |

TABLE IV. AVERAGE ERROR STANDARD DEVIATIONS OF THE VOLTAGE ANGLE

| cases | IEEE 14 Bus | | IEEE 30 Bus | |
|---|---|---|---|---|
| | ***Average Error S.D.*** | ***Percentage*** | ***Average Error S.D.*** | ***Percentage*** |
| **No PMUs** | 0.0023332 | 100% | 0.0030143 | 100% |
| **10% PMUs** | 0.0008298 | 35.56% | 0.0010244 | 33.98% |
| **20% PMUs** | 0.0001555 | 6.66% | 0.0004918 | 16.3% |
| **30% PMUs** | 0.0001311 | 5.62% | 0.0003182 | 10.56% |
| **40% PMUs** | 0.0000618 | 2.65% | 0.000317 | 10.52% |
| **Only PMUs** | 0.0000027 | 0.11% | 0.0000032 | 0.11% |

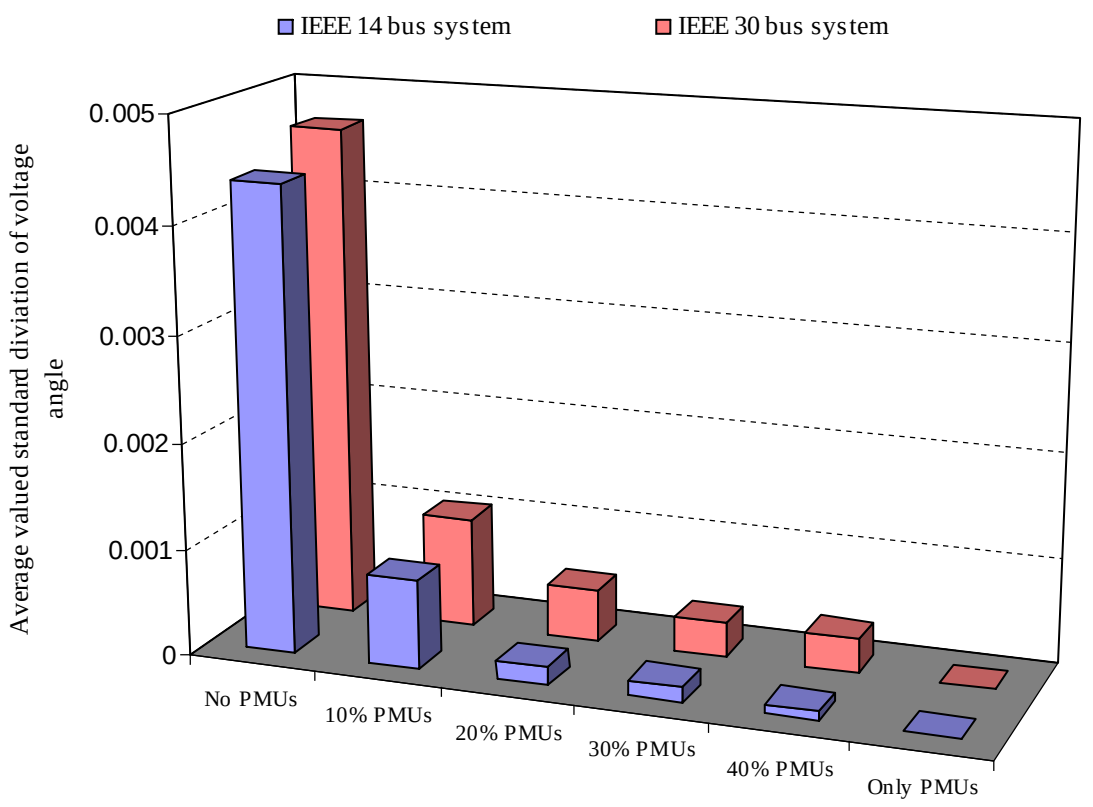

Figure 8. Average Voltage Magnitude Standard Deviation of Two Systems

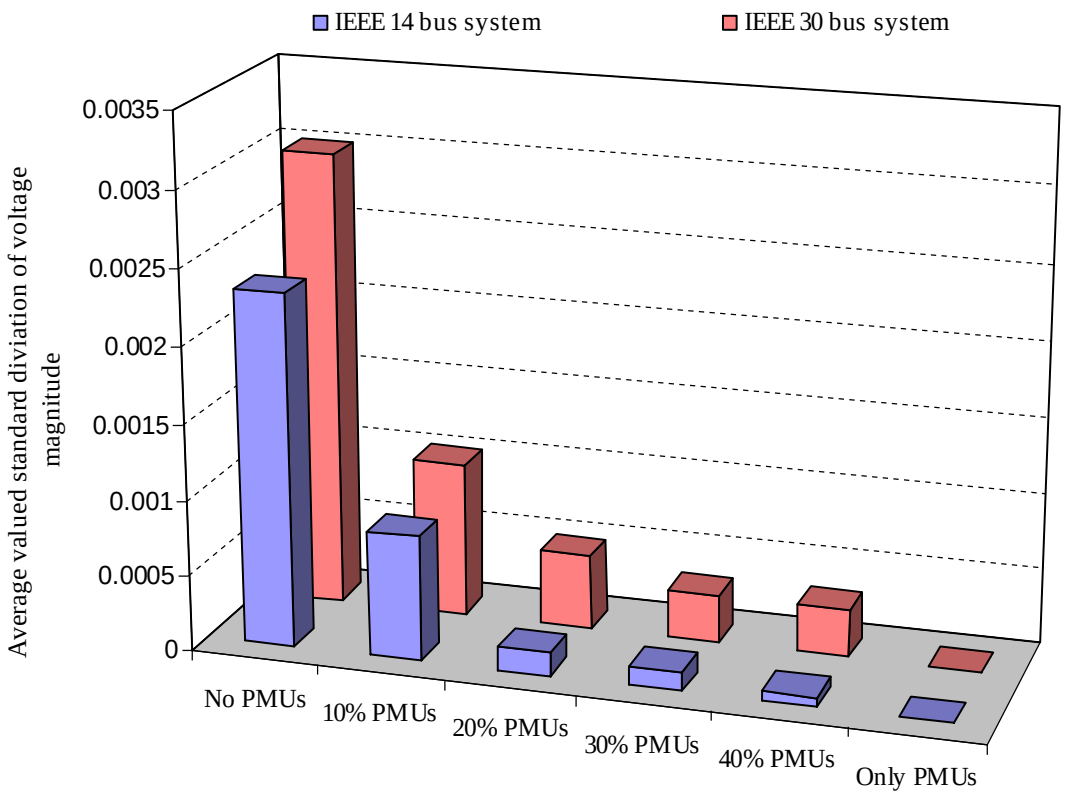

Figure 9. Average Voltage Angle Standard Deviation of Two Systems

## VII. CONCULTION

In this paper the way of incorporating the PMU data to the conventional measurements set is discussed. It is expected that those PMU measured data improve the measurement redundancy and accuracy, due to the small error standard deviations of PMU. A linear formulation of the state estimation is investigated using only PMU measured data. This linear formulation of the PMU data can produce the estimation result by a single calculation not requiring any iteration. Six cases are tested while gradually increasing the number of PMUs which are added to the measurement set by applying them on IEEE 14, 30 test systems. With the help of advanced accuracy of PMU, it was seen that the estimated accuracy is also gradually increase. One of the interesting thing is that the accuracy of estimated variables improves most effectively when the number of implemented PMUs are around ‘10%’ of the system buses.

## REFERENCES

[1] A. Abur and A. G. Exposito, Power System State Estimation, Theory and Implementation, MAECEL DEKKER,2005, pp. 9-27.

[2] Real time dynamics monitoring system [Online]. Available: http:// www.phasor-rtdms.com.

[3] A. G. Phadke and J. S. Thorp, Synchronized Phasor Measurements and Their Applications, Springer, 2008, pp. 93-97.

[4] R. Zivanovic and C. Cairns, “Implementation of PMU technology in state estimation: an overview”, IEEE AFRICON, pp. 1006-1011, 2006.

[5] Y.M. El-Fattah and M. Ribbens-Pavella, “Multi-level approach to state estimation in electric power systems, ” Proc. of the IV IFAC Symposium on Identification and System Parameter Estimation, Tbilisi, USSR, pp. 166-179, Sept. 1976.

[6] K. A. Clements, O. J. Denison, and R. J. Ringlee, “A multi-area approach to state estimation in power system networks,” in roc. IEEE Power Eng. Society Summer Meeting, San Francisco, CA, 1972, C72465-3.

[7] T. Van Cutsem, J. L. Horward, and M. Ribbens-Pavella, “A two-level static state estimator for electric power systems,” IEEE Trans. Power Apparat. Syst., vol. PAS-100, no. 8, pp. 3722–3732, Aug. 1981.

[8] L. Zhao and A. Abur, “Multiarea state estimation using synchronized phasor measurements,” IEEE Trans. Power Syst., vol. 20, no. 2, pp. 611–617, May 2005.

[9] Weiqing Jiang; Vittal, V.; Heydt, G.T, “A Distributed state estimator utilizing synchronized phasor measurements,” IEEE Tran. Power Syst., vol. 22, pp. 563 – 571, May 2007.

[10] B. Xu and A. Abur, “Optimal placement of phasor measurement units for state estimation, ” PSERC, Oct. 2005, Fin. Proj. Rep.

[11] B. Xu and A. Abur, “Observability analysis and measurement placement for systems with PMUs,” in Proc. IEEE PES Power Systems Conf. Expo., Oct. 2004, pp. 943–946.

[12] C. Rakpenthai, S. Premrudeepreechacharn, S. Uatrongjit, and N. R. Watson, “An optimal PMU placement method against measurement loss and branch outage,” IEEE Trans. Power Del., vol. 22, no. 1, pp.101–107, Jan. 2005.

[13] S. Chakrabarti and E. Kyriakides, “Optimal placement of phasor measurement units for power system observability,” IEEE Trans. Power Syst., vol. 23, no. 3, pp. 1433–1440, Aug. 2008.

[14] Y.J. Yoon, “Study of the utilization and benefits of phasor measurement units for large scale power system state estimation”, Master of Science Thesis.Texas Univ. December 2005.

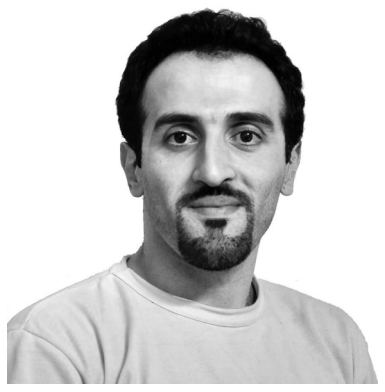

**H. Abdollahzadeh Sangrody** received the B.Sc. degree in Electrical Engineering from Zanjan University, Iran 2006. He is currently pursuing Master of Science in Electrical Engineering at Power and Water University of Technology. His research interests include Power System Observability and State Estimation in Power Systems.

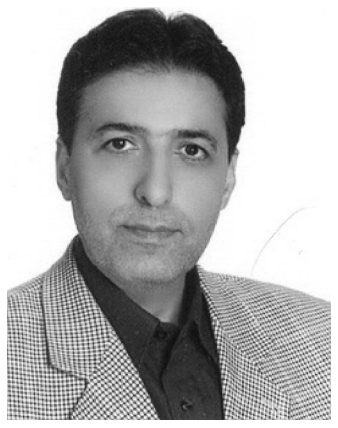

**M.T. Ameli** received B.Sc. degree in Electrical Engineering from Technical College of Osnabrueck, Germany in 1988 and M.Sc. & Ph.D. from Technical University of Berlin in 1992 & 1997. Since then He teaches and researches as associated Professor in Electrical Engineering Dept of Power & Water University of Technology in Teheran. Areas of research: Power system Simulation, Operation, Planning & Control of power system, usage of renewable Energy in Power system.

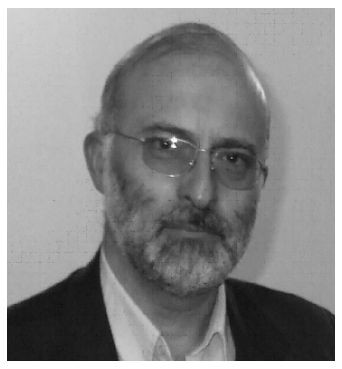

**M.R. Meshkatoddini**(SM'96) received his B.Sc. and M.Sc. degrees, in 1980 and 1990 respectively, in Electrical Engineering from Tehran Polytechnic University, Iran, and the Ph.D. degree in Electrical Engineering, from Paul Sabatier University of Toulouse, France in 1996. Since 1992, he has been a faculty member at PWUT. Dr. Meshkatoddini has more than 25 years of experience in the areas of power transformers, surge arresters, electric materials and power network transients.